# Geometric Effect of High-Resolution Electron Energy Loss Spectroscopy on the Identification of Plasmons: An Example of Graphene

Jiade Li[1,2], Zijian Lin[1,2], Guangyao Miao[1,2], Weiliang Zhong[1,2], Siwei Xue[1,2], Yi Li[1,2], Zhiyu Tao[1,2], Weihua Wang[1,3], Jiandong Guo[1,2,3*], and Xuetao Zhu[1,2,3*]

[1] Beijing National Laboratory for Condensed Matter Physics and Institute of Physics, Chinese Academy of Sciences, Beijing 100190, China

[2] School of Physical Sciences, University of Chinese Academy of Sciences, Beijing 100049, China

[3] Songshan Lake Materials Laboratory, Dongguan, Guangdong 523808, China

[*] Jiandong Guo (jdguo@iphy.ac.cn) and Xuetao Zhu (xtzhu@iphy.ac.cn).

## Abstract

High-resolution electron energy loss spectroscopy (HREELS) is one of the most powerful methods to detect the dispersion of plasmons. However, we find that in the HREELS measurement, the scattering geometric configuration will seriously affect the identification of plasmons. Here, taking graphene as an example, using the HREELS capable of two-dimensional energy-momentum mapping combined with the intensity distribution calculations, we visually display the intensity distribution of the scattering geometric factor. We demonstrate that the energy loss peaks from the scattering geometric effect may be misinterpreted as the features of an acoustic plasmon. In any HREELS measurement, it is necessary to evaluate the effect of the scattering geometry quantitatively to identify the intrinsic surface excitations.

## I. INTRODUCTION

Plasmon, the quantum of collective oscillations of the charge density in the bulk or at the surface [1, 2], not only helps us understand the many-body interactions in condensed matter physics, but also has important applications in the fields of optoelectronics [3, 4], sensors [5], and catalytic reactions [6]. High-resolution electron energy loss spectroscopy (HREELS) is one of the most powerful methods for measuring plasmons. With 1 meV energy resolution and $0 \sim 2$ Å$^{-1}$ momentum measurement range, HREELS can clearly characterize the dispersion of plasmons. Up to now, a large number of plasmons have been detected by HREELS, not only in traditional metal systems [7-9], but also in recent widely-concerned materials such as transition-metal dichalcogenides [10], topological insulators [11-14], and topological semimetals [15-18]. The discovery of these new plasmon modes not only enriches plasmon physics, but also brings broad application prospects [19, 20].

Plasmons are generally located in the dipole scattering regime of HREELS. According to the dipole scattering theory [21], the total cross section is actually the result of the competition between the scattering geometry and the intrinsic excitations of the material. In identifying the intrinsic excitations of materials, the contribution of scattering geometric effect is often ignored as a featureless background. However, the scattering geometry may show strong features, which could affect the identification of intrinsic excitations. This is especially true for semimetal systems, since the special band structure of semimetals can provide a large amount of electron-hole pair excitation channels that gives rise to the scattering background [22-24], sometimes may be extremely strong even relative to the elastic peak. The scattering background of a semimetal, such as graphene [25-27], is two orders of magnitude greater than that of insulators, semiconductors, or metals [24, 28], which makes graphene an excellent system for studying the scattering geometric effect in the HREELS experiment.

Theoretically, graphene is predicted to have plasmons with linear dispersion, called acoustic plasmons (APs) [29-32]. Experimentally, HREELS peaks on graphene showing a linear dispersion, which were considered to be the loss signals from APs, have been reported with a variety of substrates such as Cu(111) [33], Pt(111) [34], Ir(111) [35], and 6H-SiC(0001) [36]. However, the HREELS results for graphene APs have several strange behaviors, which have not been fully understood. Firstly, all the HREELS peaks of the graphene APs show asymmetric line shapes. Secondly, the reported graphene AP dispersions on different substrates, although with huge differences in the carrier density of an order of ~ 100 [35], are identical, as long as similar incident electron energies are used.

In this article, employing the HREELS with the capability of two-dimensional (2D) energy-momentum mapping (2D-HREELS) [37] combined with the intensity distribution calculations, we systematically studied the contribution of scattering geometric effect for dipole scattering by taking graphene as an example. With the visualization of the geometric effect by the 2D mapping, we show that the maximum intensity from the geometric effect is tilted with respect to the specular direction. And the tilt will produce peaks with linear "dispersion" and asymmetric shape, which do not correspond to any excitations of the sample. Especially, we point out that the AP dispersions on graphene reported by previous HREELS studies are actually the misinterpretation of the peaks caused by the geometric effect. Our work suggests that it is indispensable to deconvolute the contribution of the scattering geometric effect to identify the intrinsic excitations in a HREELS measurement.

## II. METHODS

### A. Dispersion Analyses with HREELS

The conventional HREELS collects the energy loss curves of scattered electrons at a fixed angle in each measurement [38]. The dispersion relation is achieved by rotating the monochromator, the analyzer, or the sample. We designed a system called 2D-HREELS [37], which can directly obtain a 2D energy-angle mapping simultaneously without any mechanical rotation. From a set of 2D energy-angle mapping data, one can plot the scattering intensity as a function of energy loss for a given angle, which is called the energy distribution curve at a fixed angle (aEDC). Meanwhile, one can also plot the scattering intensity as a function of angle for a given energy loss, which is called the angle distribution curve (ADC).

The scattering angle $\theta$ relative to the specular direction can be converted into momentum $q$ by following the conservation of energy and momentum

$$q = \frac{\sqrt{2m_e E_0}}{\hbar}\left[\sin\theta_0 - \sqrt{1 - \frac{E_{loss}}{E_0}}\sin(\theta_0 - \theta)\right], \quad (1)$$

where $\theta_0$ is the incident angle, $E_{loss}$ is the energy loss. One can obtain a 2D energy-momentum mapping from a 2D energy-angle mapping by Eq. (1). From a set of 2D energy-momentum mapping data, one can plot the scattering intensity as a function of energy loss for a given momentum, which is called the energy distribution curve at a fixed momentum (mEDC).

## B. Sample Preparation and Characterization

Monolayer graphene samples were prepared on a nitrogen-doped 6H-SiC(0001) substrate in a combined molecular beam epitaxy and scanning tunneling microscope/spectroscopy (STM/STS) system (Unisoku) in ultrahigh-vacuum ($\sim 1.0 \times 10^{-10}$ Torr) by solid-state graphitization [39]. A high-resolution STM image of the prepared monolayer graphene is displayed in the upper panel of Fig. 1(a), in which the honeycomb lattice can be clearly seen. In the lower panel of Fig. 1(a), the well-behaved dI/dV curves are consistent with the reported results of the monolayer graphene in the literature [40, 41].

After being transferred to the 2D-HREELS system (base pressure better than $2.0 \times 10^{-10}$ Torr), the sample was annealed at 650 °C to remove possible surface adsorptions. In this study, all the measurements in 2D-HREELS system were performed at room temperature. Surface quality and crystallographic orientation were examined by low energy electron diffraction (LEED). As shown in Fig. 1(b), the bright sharp spots in the LEED pattern demonstrate the high quality and single crystallinity of graphene on the 6H-SiC(0001) surface (Gr/SiC). Figure 1(c) shows the energy-momentum mapping of the phonons of Gr/SiC from the 2D-HREELS measurement with $E_0 = 110.2$ eV. Three phonon branches of graphene and one Fuchs-Kliewer (FK) surface phonon branch of the 6H-SiC(0001) substrate are clearly discerned, which are in good agreement with the previous experimental results [42, 43]. Figure 1(d) shows the aEDC in the specular direction with $E_0 = 20.0$ eV, consistent with the experimental results reported by Schaefer et al. [44-46]. These results demonstrate the reliability of our sample preparation and the 2D-HREELS measurements.

### III. RESULTS AND DISCUSSION

### A. Measuring Gr/SiC by 2D-HREELS

We measured the energy-loss spectra of the Gr/SiC with different incident electron energies by 2D-HREELS. Figure 2 displays the results of Gr/SiC from the 2D-HREELS measurements with $E_0 = 7.0$, 14.4, 20.0, and 110.2 eV, respectively. For comparison, the 2D energy-angle mappings, 2D energy-momentum mappings, aEDCs, and mEDCs are all plotted. Several interesting phenomena need to be noticed in the measurement results.

Firstly, all the intensity distributions of the Gr/SiC mappings are "tilted". The maximum intensity of the 2D energy-angle and energy-momentum mappings are not distributed straight along the zero-angle and momentum axis. The tilt with lower incident electron energy is more prominent than those with higher incident electron energies.

Secondly, significant tilts "produce" energy loss peaks. When the tilt of the intensity distribution is noticeable, broad peaks with dispersion will appear in the corresponding aEDCs or mEDCs (along with the sharp phonon peaks), e.g., in the cases of $E_0 = 7.0$, 14.4, and 20.0 eV as shown in Fig. 2 (a)-(l). Only when $E_0$ is large enough, e.g., the 110.2 eV case as shown in Fig. 2(m)-(p), the unnoticeable tilt does not introduce any additional peaks to the aEDCs or mEDCs.

Thirdly, there are huge differences between the aEDCs and mEDCs. Due to the angle-to-momentum conversion as expressed in Eq. (1), the degree of the tilt in the energy-momentum mapping is always larger than that in the energy-angle mapping, which makes the aEDCs and mEDCs at low incident electron energy ($E_0 \leq 20.0$ eV) show huge differences: (1) The peak height of the aEDCs is much lower than that of the mEDCs. (2) The dispersions shown in the aEDCs are slightly weaker than the mEDCs. (3) All the peak shapes are asymmetric, especially obvious in the aEDCs. (4) The peak width of the aEDCs is larger than that of the mEDCs. Noticeably, there is a large error in measuring the width of the loss peaks through the aEDCs, because the momentum corresponding to each energy point on the aEDC is different. For example, in the case with an incident electron energy of 14.4 eV [see Fig. 2(e)-(h)], the full width at half maximum (FWHM) of the peak in the aEDC at $\theta = -2.5°$ is 1.3 eV, and the momentum difference corresponding to the positions of half-peak height is 0.086 Å$^{-1}$. For comparison, the peak at the corresponding momentum on the mEDC has a FWHM of 0.6 eV. When the loss peak of an intrinsic excitation mode is measured by HREELS, the FWHM extracted by aEDC will underestimate the lifetime.

We notice from the 2D-HREELS measurement that in the 2D energy-momentum mappings, the degree of tilt of the maximum intensity varies with the incident electron energy $E_0$, and the corresponding dispersion also varies accordingly, which cannot be the intrinsic plasmon characteristics of the sample. The aEDCs and mEDCs we measured are consistent with the results of previous works [47, 48]. As we will discuss in the following, these peaks with linear dispersion, incident-energy dependence and asymmetric line shape actually originate from the geometric effects of the dipole scattering.

## B. The Dipole Scattering Background

In an HREELS measurement, the dipole scattering describes the scattering of incident electrons by a long-range electric dipolar field of the sample surface. The dipolar interaction is confined in a narrow angular range close to the specular direction, usually called the dipole scattering regime [38], in which plasmons and dipole-active phonons can be detected. A simple and useful form of the cross-section from the dipole scattering theory has been well established [21, 49, 50] and experimentally verified [51-53].

An illustration of the scattering geometry is plotted in Fig. 3(a). In the dipole scattering theory, the differential inelastic electron scattering cross-section is given by [21]

$$\frac{dS}{d\Omega} = \frac{Am_e|R|^2}{\pi^2}|\Delta|^2 \times G(\theta_0, \theta, \phi, \theta_E), \quad (2)$$

with

$$G(\theta_0, \theta, \phi, \theta_E) = \frac{(\theta \cos\phi \cos\theta_0 - \theta_E \sin\theta_0)^2 + \theta^2 \sin^2\phi}{E_0 \cos\theta_0 (\theta^2 + \theta_E^2)^2},$$

where $A$ is the surface area, $R$ is the probability amplitude that describes the specular reflection of the incident and scattered electrons from the surface. $|\Delta|^2$ is the interaction term between the incident electrons and the sample, reflecting the intrinsic electronic response of the material, which can be treated as angular independent. $\phi$ is the out-of-plane scattering angle. $\theta_E = E_{loss}/2E_0$ is the characteristic angle (see Ref. [54] for details). The angular-dependent term in Eq. (2), $G(\theta_0, \theta, \phi, \theta_E)$, is only related to the scattering geometric configuration, which is defined as the scattering geometric factor (SGF). In an HREELS measurement, the intrinsic excitations of the material and the SGF together contribute to the intensity distribution of the dipole scattering.

Figure 3(b) shows the intensity distribution of the SGF with $\theta_0 = 45°$, $\theta_E = 0.1$ rad, and $\phi = 0$. Clearly, the intensity distribution in the specular direction is asymmetric. There are two scattering lobes: one is toward the sample normal mainly located in the forward scattering region, and the other is toward the sample surface mainly located in the backward scattering region. The forward scattering lobe is much stronger than the backward scattering lobe — in an HREELS measurement, the dipole scattering intensity mainly comes from the forward scattering, which is distributed within a small angle range, with its maximum value deviating from the specular direction, as clearly illustrated in the ADC in Fig. 3(c).

At a fixed incident angle $\theta_0$, the location of the maximum scattering intensity and the width of the forward scattering lobe are modulated by $\theta_E$ [52, 53]. When the magnitude of $\theta_E$ is large enough (for example, $\geq 0.02$), the impact by the SGF on the HREELS spectra will be manifested as "ghost" loss peak which is not due to any particular discrete loss channel but is a consequence of the intrinsic kinematics of the electron energy-loss probability function. This phenomenon has been discovered theoretically and verified experimentally in previous works [24, 28]. This shows that the SGF is not a featureless background, but has obvious dispersion feature. Especially in the case of measuring the dispersion of low-dimensional plasmons ($\sim$ 1 eV) at low incident electron energy ($E_0 \leq 20$ eV), i.e., $\theta_E = E_{loss}/2E_0 \geq 0.025$, the SGF is too strong to be neglected. For example, the slope of the dispersion caused by the SGF is very close to that of the AP [33-36, 55-61]. However, the effects of the SGF are difficult to be distinguished directly from APs when measured with the conventional HREELS due to the lack of mapping along the angular direction.

## C. Identifying the SGF Background

In this section, we will demonstrate that the strange intensity distribution in the 2D mappings with various incident electron energies measured in the current work can be completely reproduced by SGF without relying on any parameters of the sample.

Figure 4 shows the calculated ADCs (black solid lines) based only on the SGF via Eq. (2), together with the corresponding experimental results (red open circles) for the energy loss of 0.5, 1.0, and 1.5 eV extracted from the 2D energy-angle mapping with $E_0$ = 110.2 and 14.4 eV [ADC along the white dashed lines in Fig. 2(m) and 2(e)]. For the case of $E_0$ = 110.2 eV, the deviation of the maximum intensity from the specular direction is not obvious. However, for the case of $E_0$ = 14.4 eV, the maximum intensity obviously deviates from 0° at the same energy loss. As discussed in the previous section, the degree of tilt in the energy-angle mappings is modulated by $\theta_E$ ($\theta_E = E_{loss}/2E_0$). At a fixed $E_0$, $\theta_E$ becomes larger with the $E_{loss}$ increasing, and accordingly, the angle (negative) where the maximum of the scattered signal is located increases [Fig. 4(d) to (f) for example]. This leads to the tilt of the maximum intensity in the energy-angle mappings. On the contrary, at a fixed $E_{loss}$, when $E_0$ increases, $\theta_E$ becomes smaller, and the resulting degree of the tilt will be smaller [take the comparison of Fig. 4(a)-(c) and (d)-(e) as an example]. This is the reason why the degree of tilt in the energy-angle mappings decreases as $E_0$ increases. These characteristics can be derived from Eq. (2) and have been proved by previous theoretical and experimental studies [52, 53].

Next, we will demonstrate that the aEDC peaks with asymmetric shapes and linear dispersion are actually produced by the SGF. In order to illustrate the origin of the asymmetric peak shape, in Fig. 4(g), we plot the ADCs of the calculated results ($E_0 = 14.4$ eV) with energy loss at 0.25, 1.00, and 1.50 eV as an example. When $\theta$ is within the shadow range of Fig. 4(g), the intensity of the 1.00 eV ADC is larger than the 0.25 and 1.50 eV ADCs. In this case, the peak caused by the SGF will appear as an off-specular peak in the aEDC. According to the calculation result, this peak shows an asymmetric shape [see inset in Fig. 4(g)]. The experimental aEDC at $\theta = $ -2.5° is shown in Fig. 4(h). The measured data can be fully reproduced by our calculation combined with the elastic tail and FK phonon of 6H-SiC(0001). It is worth mentioning that the asymmetric peak shapes observed in graphene on Cu (111) [33], Pt (111) [34, 62], and Ir (111) [35, 48] substrates, were previously attributed to multiple plasmon excitations [35, 48]. As demonstrated above, however, the asymmetric line shape is actually introduced by the SGF. Furthermore, the calculated dispersion of the asymmetric SGF peaks is shown together with the experiment results in Fig. 4(i). The linear dispersion of the observed peak is completely described by the model of the SGF without any sample parameters.

## D. Acoustic Plasmon vs. Scattering Geometric Factor

The above discussions suggest that the APs of graphene on different substrates reported previously can be attributed to SGF without any controversy remained. Figure 5(a) shows the dispersion data of the AP in graphene on various substrates extracted from previous works [33-35, 47, 48] superimposed on our 2D energy-momentum mapping. Obviously, these dispersion data are all located in the vicinity of the maximum scattering intensity described by the SGF. Using the aEDCs shown in Fig. 2, we can obtain the energy-momentum points contributed only by the SGF. In Fig. 5(b), a careful comparison of the SGF dispersions measured by 2D-HREELS in the current work and the AP dispersions reported previously indicate substantially similar slopes when the incident electron energies are similar. What is particularly noteworthy is that the slope decreases with the decreases of the incident electron energy, regardless of the substrate under graphene. Carrier densities of graphene on different substrates differ by two orders of magnitude [35]. The fact that the dispersion slope only changes with the incident electron energy but is independent on the carrier density suggests that the measured peaks originate from the SGF only and have nothing to do with the plasmon excitations of graphene.

In addition, we have also noticed that the graphene APs reported in previous experiments were observed only when the incident electron energy was not higher than 20 eV. This can be easily explained by the fact that the tilt from the SGF is too small to produce a peak on the aEDC, when the incident electron energy is high. For example, no loss signal appears on the aEDC with $E_0 = 110.2$ eV [Fig. 2(m)-(p)].

In this context, it will be necessary to exclude the contribution of the SGF before identifying the intrinsic excitations from the HREELS measurement. Here we give two suggestions for identifying intrinsic excitations: (1) If the intensity of the specular direction does not decay continuously with the energy loss, but a loss peak appears, then this loss peak must correspond to an intrinsic excitation, but not the SGF. This is clear for excitations with energy gaps at zero momentum, such as optical phonons or three-dimensional plasmons. (2) If the intensity decays continuously with the energy loss, and the maximum intensity tilts along the zero angle, then the measurements with different incident electron energies must be performed. If the degree of tilt changes with the incident electron energy, then the signal originates from the SGF, otherwise, the signal originates from the intrinsic excitations of the sample. These are not complicated if the 2D-HREELS mappings are obtained. But for the conventional HREELS, if there is a single loss peak in the aEDC measured from the off-specular direction, the calculation of SGF by Eq. (2) must be applied to identify the origin of the peak.

## IV. SUMMARY

In this article, we systematically studied the geometric distribution of the dipole scattering in HREELS measurements. Our 2D mapping provides a direct picture to visualize the SGF of the dipole scattering. We found that the SGF in dipole scattering will "produce" peaks with linear dispersion and asymmetric line shape in the forward scattering, not corresponding to any intrinsic excitations of the sample. The slope of the dispersion decreases with the decrease of the incident electron energy. Using graphene as a prototypical example, we found that the measured scattering distribution is in full compliance with the description of the SGF. The peaks from the SGF may be misinterpreted as the features of the graphene AP. Here, it is worth pointing out that the impact of SGF is not limited to identifying APs. Actually, all excitations in HREELS measurement, especially in the cases with low incident electron energy, may be affected by the background from the SGF. We emphasize that careful deconvolution of the contribution from the SGF is indispensable for the identification of any intrinsic excitations from HREELS measurement.

**Declaration of Competing Interest**

The authors declare that they have no known competing financial interests or personal relationships that could have appeared to influence the work reported in this paper.

**Acknowledgement**

This work was supported by the National Key R&D Program of China (Nos. 2021YFA1400200 and 2017YFA0303600), the National Natural Science Foundation of China (Nos. 11874404, 11974399, 11974402, and 11634016), and the Strategic Priority Research Program of Chinese Academy of Sciences (No. XDB33000000). X.Z. was partially supported by the Youth Innovation Promotion Association of Chinese Academy of Sciences.

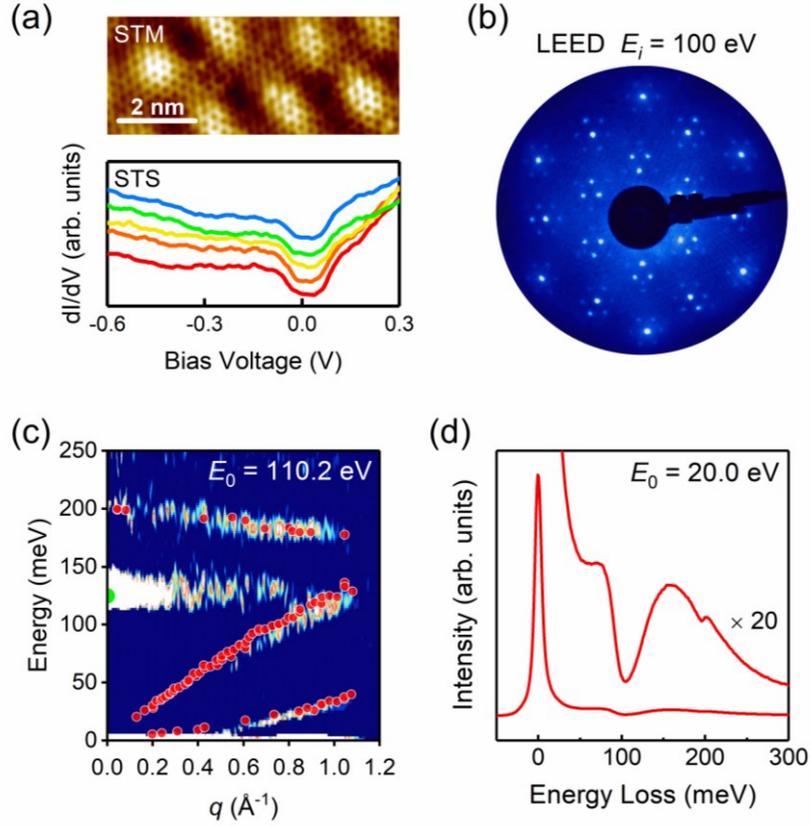

**FIG. 1. Characterization of Gr/SiC.** (a) Atomic resolved STM image (-100 mV, 3 nA) (upper) and STS spectra at different terraces (lower), at a sample temperature of 77 K. (b) LEED pattern with incident electron energy of 100 eV. (c) The second derivative image of the phonon spectrum obtained by 2D-HREELS with $E_0 = 110.2$ eV. Red and green circles mark previous HREELS results extracted from Ref. [42] and [43], respectively. $E_0 = 110.2$ eV. (d) Energy loss spectra in the specular direction with $E_0 = 20.0$ eV.

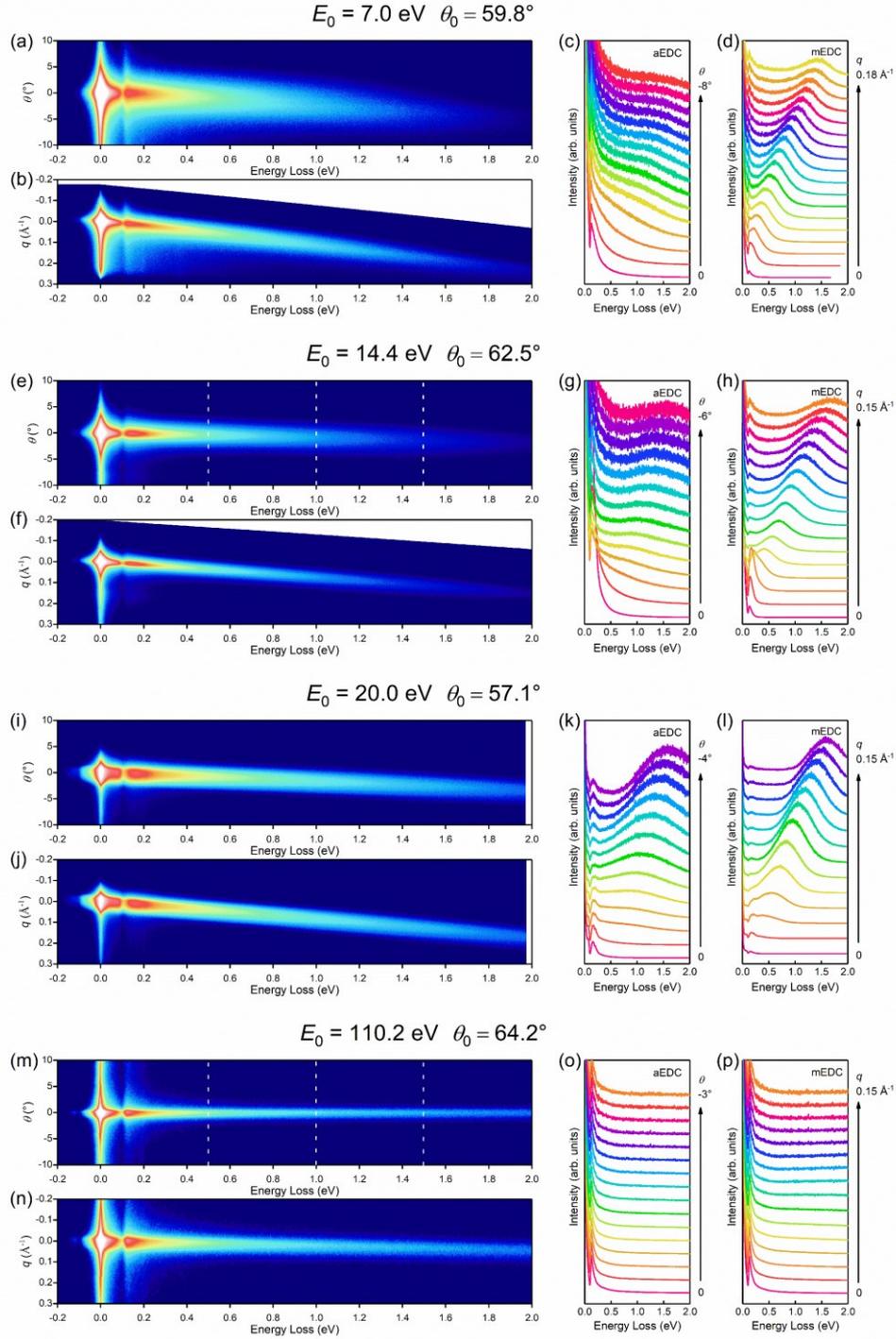

**FIG. 2. Results of the 2D-HREELS measurement.** (a) 2D energy-angle mapping, (b) 2D energy-momentum mapping, (c) aEDC stack, and (d) mEDC stack with $E_0$ = 7.0 eV. (e)-(h), (i)-(l), and (m)-(p) The corresponding results with $E_0$ = 14.4, 20.0, and 110.2 eV, respectively.

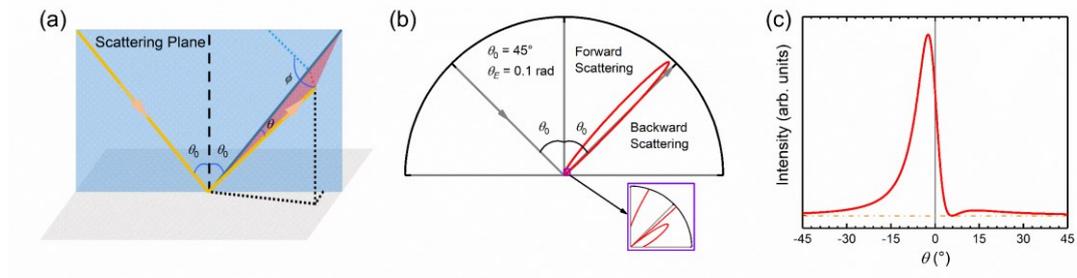

**FIG. 3. Scattering geometry of HREELS and the intensity distribution of the dipole scattering.** (a) The scattering geometry of a typical HREELS measurement. (b) The intensity distribution in a polar coordinate calculated with Eq. (2). Inset: zoom-in of the purple square to show the details of the backward scattering. (c) The intensity distribution as a function of $\theta$, with $\theta_0 = 45°$, $\theta_E = 0.1$ rad, $\phi = 0$.

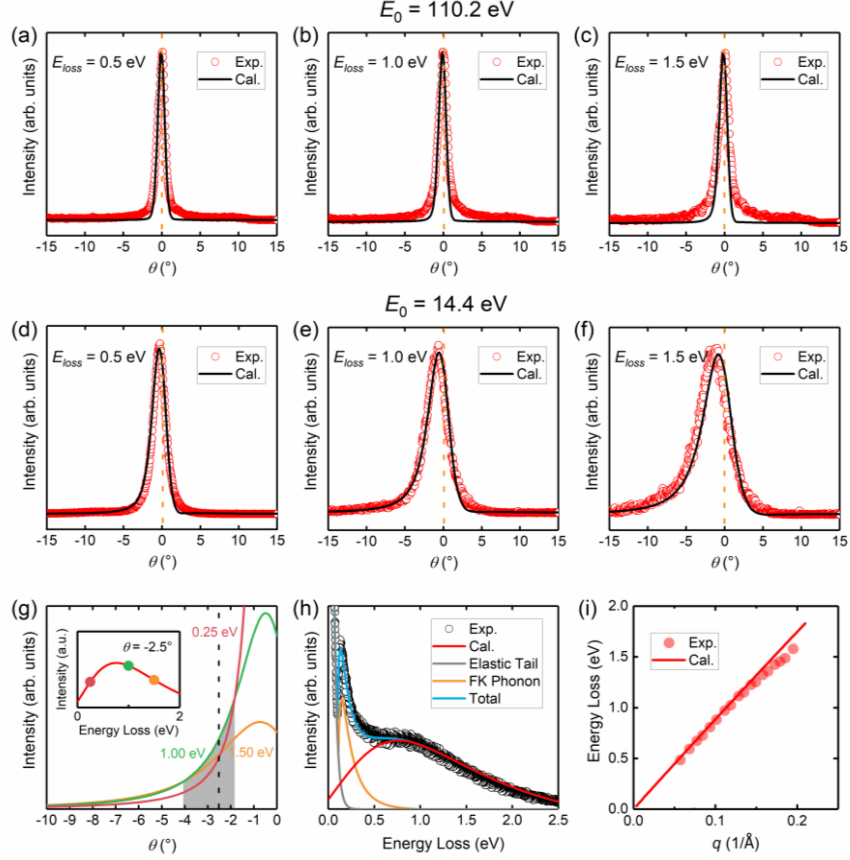

**FIG. 4. Comparison of the calculated and measured results.** (a), (b) and (c) the ADCs correspond to energy loss of 0.5, 1.0, and 1.5 eV with $E_0$ = 110.2 eV, respectively. (d), (e) and (f) correspond to the results with $E_0$ = 14.4 eV. The experimental data are extracted from the white dashed lines from Fig. 2(e) and 2(m). (g) Calculated ADCs with energy loss at 0.25, 1.00, and 1.50 eV (corresponding to the claret, green, and orange solid lines, respectively). Insert: calculated aEDC at $\theta$ = -2.5°. Solid circles correspond to the position of the intersections of the dashed line with the ADCs. (h) Experimental aEDC at $\theta$ = -2.5° extracted from Fig. 2(e). (i) Calculated dispersion from the SGF (solid red line) and experimental results (solid red circles). (g)-(i) corresponds to $E_0$ = 14.4 eV.

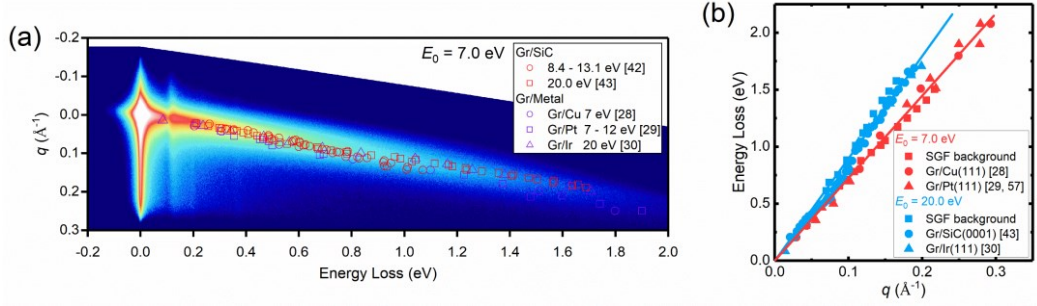

**FIG. 5. Comparison of the SGF lobe with the AP dispersion data of graphene.** (a) Dispersion data extracted from previous works [33-35, 47, 48] superimposed on 2D energy-momentum mapping with $E_0$ = 7.0 eV. (b) A careful comparison of the dispersion slope extracted from aEDCs. The AP dispersion data is extracted from Ref. [33-35, 48, 62]. The solid lines are the guide lines.